\documentclass[11pt,a4paper]{article}
\pdfoutput=1
\usepackage{empheq}
\usepackage{cases}
\usepackage{a4wide}
\usepackage{amsfonts}
\usepackage{amssymb}
\usepackage{amsmath,bm}
\usepackage{amsmath}
\usepackage{graphicx}
\usepackage{mathtools}
\usepackage{booktabs}
\usepackage{array}
\usepackage{color}
\usepackage{ulem}
\usepackage[small,bf]{caption}
\usepackage{cite}
\usepackage[usenames,dvipsnames]{xcolor}
\usepackage{subfigure}
\usepackage{verbatim}

\allowdisplaybreaks

\numberwithin{equation}{section}

\newcounter{mysubequation}[equation]

\DeclarePairedDelimiterX\braket[2]{\langle}{\rangle}{#1 \delimsize\vert #2}

\begin{document}
\begin{titlepage}

\begin{center}
{
\bf\LARGE Synchronization Induced by Ultralight Dark Matter
}
\\[8mm]
Chrisna~Setyo~Nugroho\footnote[1]{setyo13nugros@gmail.com}  
\\[1mm]
\end{center}
\vspace*{0.50cm}

\centerline{\it Department of Physics, National Taiwan Normal University, Taipei 116, Taiwan}
\vspace*{1.20cm}

\begin{abstract}
\noindent
We study the possibility of ultralight bosonic dark matter (UlDM) to induce a synchronized behaviour on randomly
evolving oscillators. We introduce a new interaction between UlDM and the oscillators and further
demonstrate the occurence of the phase transition from disordered oscillators state into ordered
state. For UlDM within the mass range of $10^{-14}\,\text{eV}/c^{2} \leq m_{\phi} \leq 1 \,\text{eV}/c^{2}$ as
well as the white noise strength lower than 0.01 Hz, the synchronization of randomly
distributed oscillators with the mean angular frequency $1$ Hz  and the corresponding standard
deviation 0.1 Hz occurs when the coupling strength between UlDM and oscillators lies within $10 \geq g_{\omega} \geq 10^{-13}$. This offers an alternative method to unveil the nature of UlDM with
respect to the established experiments.     
\end{abstract}

\end{titlepage}
\setcounter{footnote}{0}

\section{Introduction}

Many phenomena observed in nature exhibit collective behaviour such as the synchronized motion of
oscillators that evolves into a common frequency in spite of the variation in the natural frequencies
of each oscillators.  To name a few, a swarm of fireflies flashes in unison \cite{Buck:1976bjb}, the synchronized chirping
of crickets \cite{Walker:1969wlk}, the entrainment of microwave oscillators \cite{York:1991ray}, and Josephson junction which shows  superconducting behaviour \cite{Wiesenfeld:1996dtz}. The transition of initially disordered state into coherence state has brought many physicists
to study these emergent phenomena, in particluar those who work on complex system and non-linear
dynamics. 

On the other hand, the matter content of our universe is dominated by non-luminous matter dubbed as dark matter (DM).
The existence of dark matter comes
solely from its gravitational force as implied from various astrophysical observations such as
galactic rotational curve and bullet cluster. Other properties such as its mass, spin, as well as its
minuscule electric charge remain under active investigation. The mass range of DM in particular, spans
a wide range of magnitudes from $10^{-22} \, \text{eV}/c^{2}$ to the mass scale of primordial blackhole
(pBH) which calls for different detection strategies. As one of the most popular DM candidates, the null
result from  direct searches \cite{XENON:2018voc}, indirect searches \cite{Fermi-LAT:2016uux}, collider searches \cite{ATLAS:2017bfj} as well as unconventional searches \cite{Tsuchida:2019hhc,Lee:2020dcd,Chen:2021apc,Ismail:2022ukp,Lee:2022tsw,Stadnik:2014tta,Arvanitaki:2015iga,Hall:2016usm,Chen:2022abz,Nugroho:2023cun,Nugroho:2024ltb,Chen:2025tlx} of
weakly interacting massive particle (WIMP) within the mass range from 1 GeV$/c^{2}$ to 100 TeV$/c^{2}$ has shifted the interest of physicsts towards the lighter DM
mass. 

Ultralight dark matter with mass below 1 eV$/c^{2}$ has
gained a significant attention in the past few years. In contrast of the particle nature of WIMP \cite{Camargo:2019ukv,Belyaev:2016lok,Alves:2016bib,Chen:2019pnt,Dirgantara:2020lqy,Chen:2024jbr,Ismail:2024bjw}, UlDM
behaves like classical coherent waves thanks to its high occupation number mode \cite{Kimball:2023vxk}. Furthermore, the wavy
UlDM oscillates at its Compton frequency $\omega_{\phi} = m_{\phi} \,c^{2}/\hbar$ with $m_{\phi}$
denoting its mass. Many experiments has been carried out to detect UlDM such as employing microwave
cavity \cite{Hagmann:1990tj,DePanfilis:1987dk}, magnetic resonance search \cite{Budker:2013hfa}, dark matter radio \cite{Chaudhuri:2014dla}, lab searches for exotic spin-dependent
interactions \cite{ARIADNE:2017tdd}, light shining through wall experiments \cite{Ehret:2010mh,Ortiz:2020tgs}, as well as global quantum sensor networks \cite{Afach:2018eze}. In
this paper, we study the synchronized behaviour of oscillators induced by the wavy UlDM. We show that
the wave nature of UlDM would transform the initially disordered state of oscillators into coherent
state. We estimate the sensitivity of a particular dsitribution of oscillators to be entrained by UlDM.         
 
This paper is structured as follows: We discuss the interaction between oscillators and UlDM in 
Section \ref{sec:interaction}. We further investigate the minimum coupling strength required to synchronize the oscillators for a given white noise in
Section~\ref{sec:sensitivity}. In the same section, we also discuss the projected sensitivity of our study.
Finally, the summary and conclusion of our study are presented in Section~\ref{sec:Summary}.

\section{Interaction between Ultralight Dark Matter and Oscillators}
\label{sec:interaction}

Synchronization phenomenon is well described by set of coupled oscillators called the
Kuramoto model \cite{Kuramoto:1975kra,Kuramoto:1984kur} 
\begin{align}
\label{eq:kuraEq}
\dot{\theta}_{i} = \omega_{i} + \frac{K}{N}\, \sum^{N}_{j=1} \text{sin} (\theta_{j} - \theta_{i})\,,
\end{align}
where $\theta_{i}$, $\omega_{i}$, and $K$ denote the phase of the i-th oscillator, the natural
frequency of the i-th oscillator as well as its corresponding coupling with respect to other
oscillators, respectively. The appearance of $N$ in the denominator guarantees that the model is well behaved in the
limit of $N \rightarrow \infty$. When $K > 0$, the interactions between oscillators are attractive. To
see this, one may consider two oscillators $i$ and $j$, with oscillator $j$ evolves slightly ahead of oscillator $i$.
Consequently, the phase difference between them $\theta_{j} - \theta_{i}$ will be positive\footnote{Assuming that $\theta_{i},\,\theta_{j}$, and their difference is located within the interval $0 \leq \theta \leq \pi/2$.}. Since $K > 0$, this interaction gives a positive contribution to $\dot{\theta}_{i}$. On the other hand, $\dot{\theta}_{j}$ would move slower due to this interaction which make these two oscillators eventually
synchronize.
\begin{figure}
	\centering
	\includegraphics[width=0.8\textwidth]{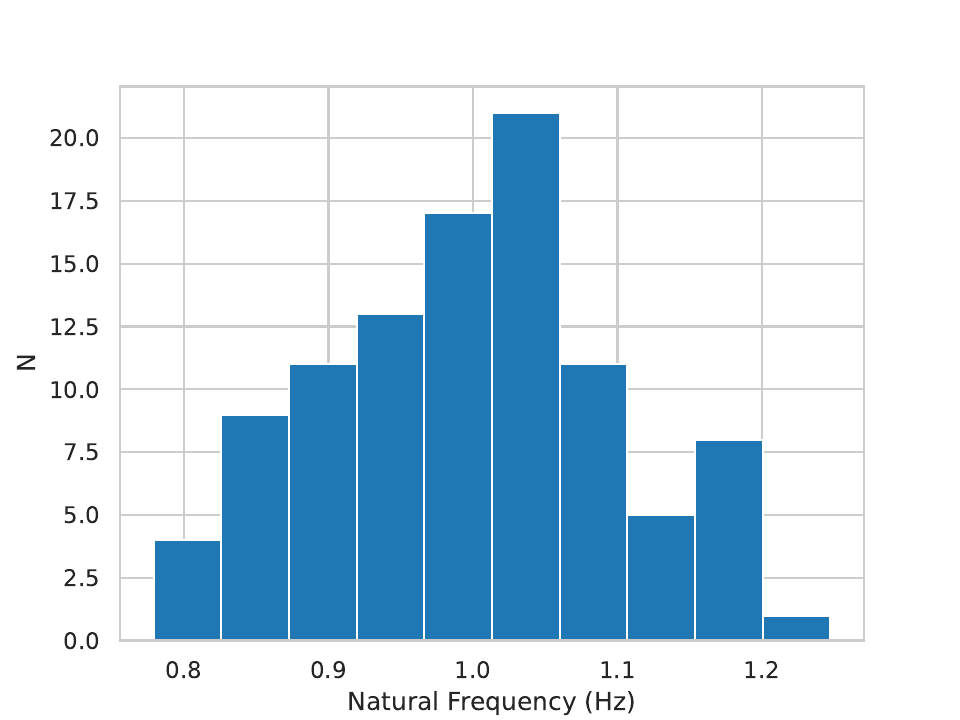}
	\caption{The distribution of 100 oscillators with the mean frequency $\mu = 1$ Hz and standard deviation $\sigma=$ 0.1 Hz.}
	\label{fig:dist}
\end{figure}
To understand the dynamics of these oscillators, it is useful to label each phase of the
oscillators $\theta_{j}$ as a point at two dimensional plane $(\text{cos} \,\theta_{j}, \text{sin}\,\theta_{j})$ located on the unit circle. Moreover, we may parameterize such point as complex number $\text{cos}\,\theta_{j} + i\, \text{sin}\, \theta_{j} = e^{i\,\theta_{j}}$. Since each oscillator corresponds to different $e^{i\,\theta_{j}}$, one may take the average value as \cite{Kuramoto:1984kur,Strogatz:2000str}
\begin{align}
\label{eq:Cr}
r\, e^{i\,\Psi} = \frac{1}{N}\, \sum^{N}_{j=1} \, e^{i\,\theta_{j}}\,.
\end{align}
The quantity on the left hand side of Eq.\eqref{eq:Cr} is called the complex order parameter ($0 \leq r \leq 1$) which can be regarded as the centroid of the oscillators $e^{i\,\theta_{j}}$ located on the
unit circle. The magnitude of the complex order parameter or the order parameter $r$ determines the degree of phase coherence
or order in the population of oscillators. To see this, one can take the extreme limit where all
oscillators have the same phase $e^{i\, \alpha}$. In this case, one may verify that $r = 1$ and $\Psi = \alpha + 2 \pi n$ with $n$ denoting the integer. In contrast, when the oscillators are spread with $\theta_{j} = 2\pi j /N$, one can check that $r = 0$. The angle $\Psi$ is interpreted as the measure of
the average phase of the oscillators. In terms of complex order parameter, the Kuramoto equation can be written as \cite{Kuramoto:1984kur,Strogatz:2000str}
\begin{align}
\label{eq:RkuraEq}
\dot{\theta}_{i} = \omega_{i} + K\,r\, \text{sin}(\Psi - \theta_{i})\,.
\end{align}
Here, the problem originally viewed as one oscillator interacts with every other oscillators can be reinterpreted as one oscillator coupled to the mean field produce by all oscillators encoded in $r$ and $\Psi$.
\begin{figure}
	\centering
	\includegraphics[width=1.0\textwidth]{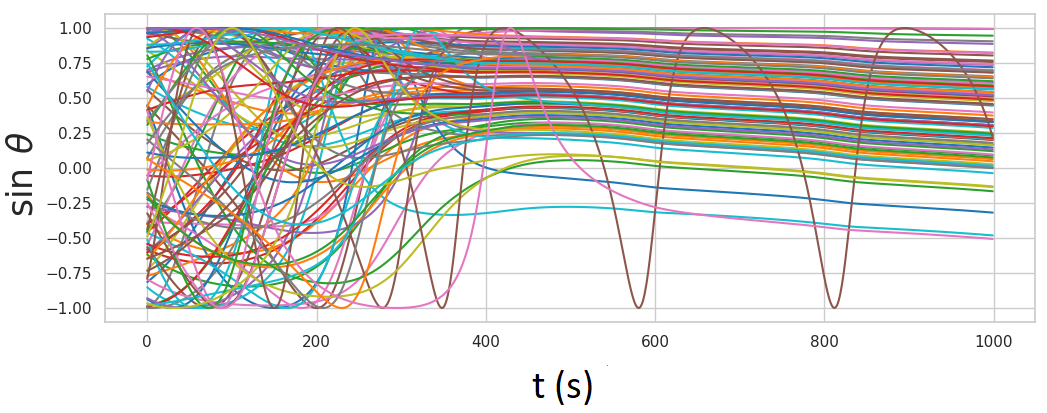}
	\caption{The phase evolution of 100 oscillators with respect to time.}
	\label{fig:phaseVSt}
\end{figure}
Ultralight dark matter may induce synchronization on independent non-interacting oscillators thanks to
its  wave nature. To understand the dynamics of oscillators in the presence of UlDM, one needs to model
the coupling $K$ between the oscillators and UlDM. Since UlDM oscillates at the Compton frequency $\omega_{\phi}$ and it has the same dimension with $K$ appearing in Kuramoto model, it is natural to take 
\begin{align}
\label{eq:KUldm}
K = g_{\omega} \frac{m_{\phi} \, c^{2}}{\hbar}\,,
\end{align}
where $g_{\omega}$ stands for the dimensionless frequency coupling between oscillators and UlDM. Thus,      the time evolution of phase oscillator $\theta_{i}$ with natural frequency $\omega_{i}$ in the background of UlDM is given by 
\begin{align}
\label{eq:EqKuraDM}
\dot{\theta}_{i} = \omega_{i} + g_{\omega} \frac{m_{\phi} \, c^{2}}{\hbar}\,r\, \text{sin}(\Psi - \theta_{i})\,.
\end{align}

To demonstrate the synchronization induced by UlDM, we consider 100 oscillators which have natural frequencies distributed according to Gaussian function
\begin{align}
\label{eq:Gauss}
G(\omega) = \frac{1}{\sigma\, \sqrt{2\pi}} e^{-\frac{(\omega-\mu)^{2}}{2\sigma^{2}}}\,,
\end{align}
with the corresponding mean frequency $\mu = 1 $ Hz and the standard deviation $\sigma = 0.1$ Hz as shown Fig.~\ref{fig:dist}. Furthermore, we numerically solve Eq.~\eqref{eq:EqKuraDM} by using a \texttt{python} package \cite{Fabrizio:2019fbr} where we set the frequency coupling $g_{\omega} = 3 \times 10^{-14}$ and UlDM mass $m_{\phi} = 0.01\, \text{eV}/c^{2}$.  
 
\begin{figure}
	\centering
	\includegraphics[width=1.0\textwidth]{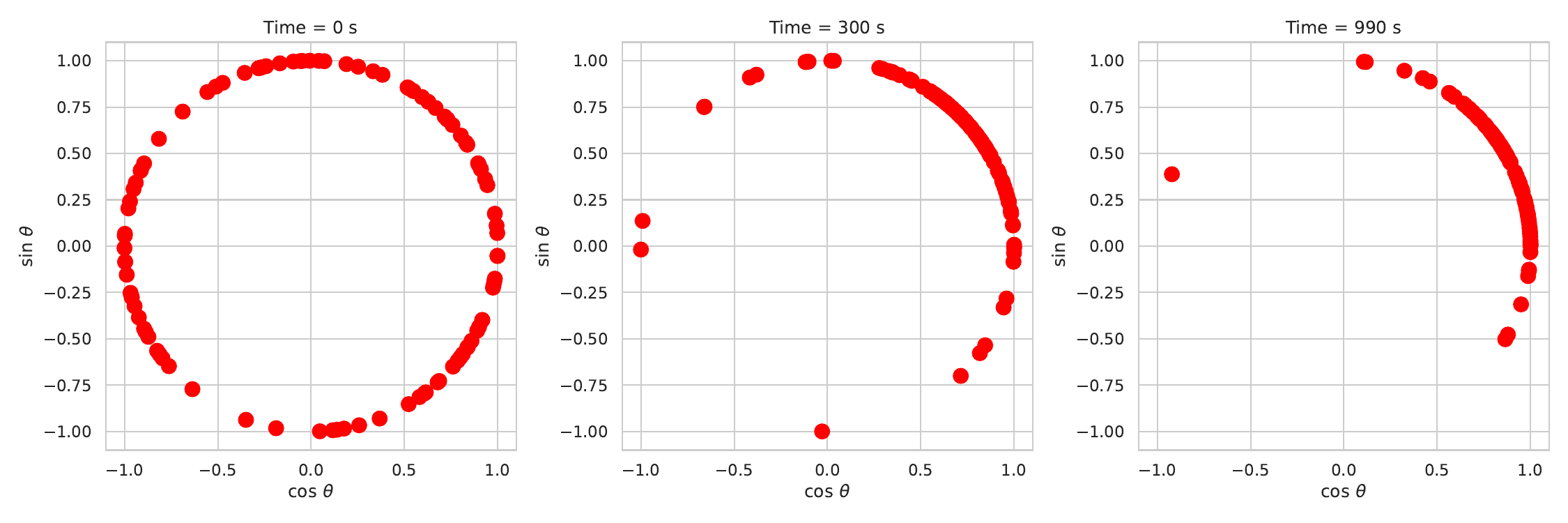}
	\caption{The snapshot of 100 oscillators in $(\text{cos}\,\theta, \text{sin}\,\theta)$ plane at 3 different time.}
	\label{fig:test}
\end{figure}

The phase evolution of 100 oscillators as function of time is displayed in Fig.~\ref{fig:phaseVSt}. There, one can see that all oscillators evolve according to their natural frequencies until t = 300
s. After 300 s, the majority of the oscillators oscillate with constant phase. This behavior can be seen clearly when
we plot the oscillators evolution in $(\text{cos}\,\theta, \text{sin}\,\theta)$ plane as shown in Fig.~\ref{fig:test}.   

Initially, all oscillators are randomly distibuted according to their natural frequencies as can be seen
in the left panel of Fig.~\ref{fig:test}. Eventually, the oscillators start to move in unison as
depicted in the middle panel of Fig.\ref{fig:test} where the clustering of the population of the
oscillators start to form. Finally, at t = 990 s the clustering become apparent with most of the oscillators experience synchronization. 
\begin{figure}
	\centering
	\includegraphics[width=0.7\textwidth]{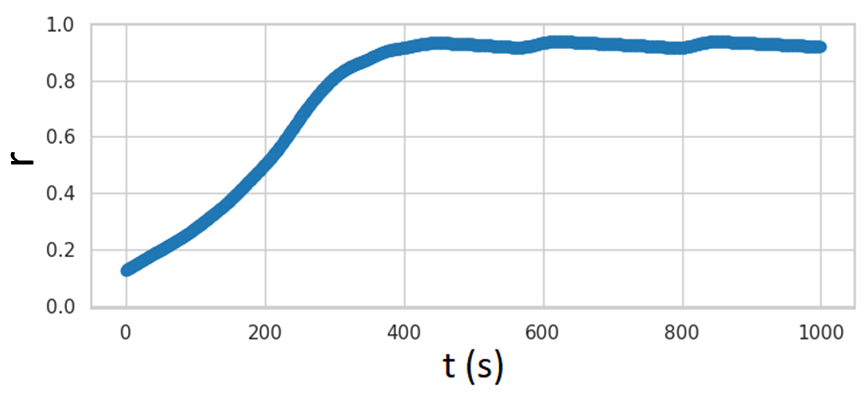}
	\caption{Time evolution of the order parameter $r$ for 100 oscilators under consideration.}
	\label{fig:rVST}
\end{figure}

In the language of the order parameter $r$, the degree of coherence of 100 oscillators under
consideration is depicted in Fig. \ref{fig:rVST}. There, the oscillators exhibit disordered state
initially which is obvious since each of them evolves according to its natural frequency. Due to their
interaction with UlDM, each oscillators start to move in unison as can be inferred from the increase of
the order parameter in Fig.~\ref{fig:rVST}. As time passes, more oscillators join the synchronized
group and the system becomes more ordered. The wiggle observed at the saturation value of $r$ stems from the fact that we have finite size of oscillators ($N = 100$) with fluctuation size $\Delta r \approx 1/\sqrt{N}$ \cite{Strogatz:2000str}.  
\begin{figure}
	\centering
	\includegraphics[width=0.62\textwidth]{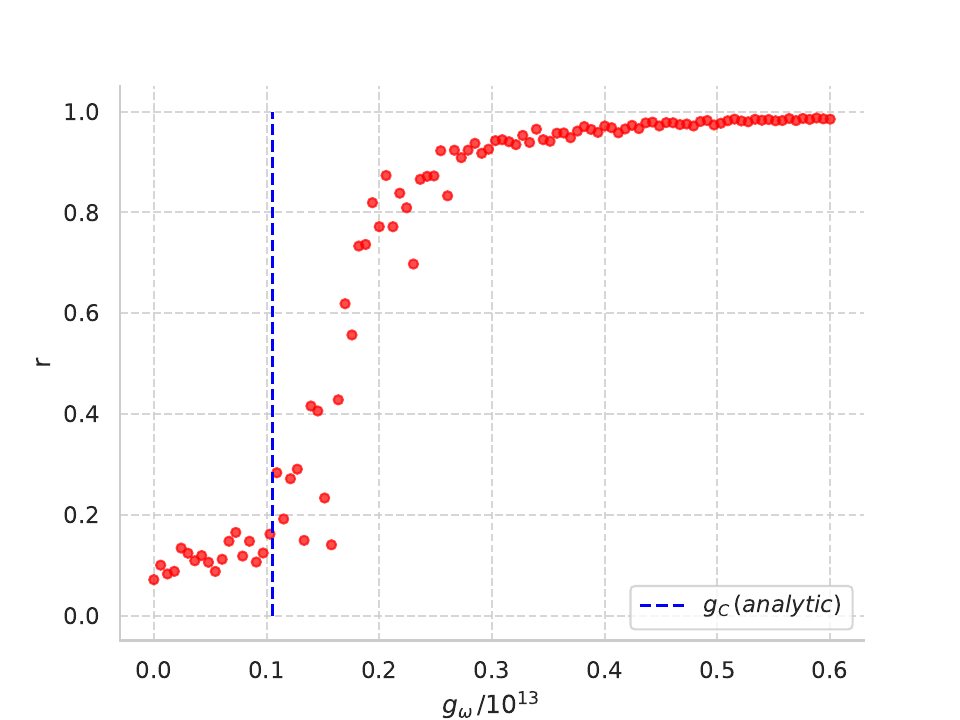}
	\caption{The behaviour of $r$ as function of $g_{\omega}$. It indicates the phase transition with critical coupling $g_{C}$ as depicted by vertical dashed line.}
	\label{fig:rVSK}
\end{figure}  

In Fig.~\ref{fig:rVSK}, we
show the behaviour of $r$ with respect to $g_{\omega}$ for $m_{\phi} = 0.01\, \text{eV}/c^{2}$. We see
that $r$ remains fixed as $g_{\omega}$ increases until a particular value of $g_{\omega} = g_{C}$. When $g_{\omega} > g_{C}$, the positive change of $g_{\omega}$ enhances the value of $r$ before it reaches the saturation point. This demonstrates the existence of the phase transition of the oscillators from disordered state to ordered state with the value of critical coupling\footnote{In terms of the original coupling $K$, the critical coupling is given by $K_{C} = \sigma \, \sqrt{\frac{8}{\pi}}$ \cite{Strogatz:2000str}.}
\begin{align}
\label{eq:gC}
g_{C} = \frac{\hbar\,\sigma}{m_{\phi}\,c^{2}}\, \sqrt{\frac{8}{\pi}}\,,
\end{align}
where $\sigma$ is the standard deviation of Gaussian dsitribution in Eq.~\eqref{eq:Gauss}. For $\sigma = 0.1 $ Hz, the value of the critical coupling is $g_{C} \approx 1.1\times 10^{-14}$.

\section{Projected Sensitivity}
\label{sec:sensitivity}

For a given distribution of oscillators at a fixed UlDM mass, one can determine the critical coupling $g_{C}$ from Eq.\eqref{eq:gC}. Thus, one may estimate the sensitivity of a particular oscillator
population to be entrained by UlDM. However, the original Kuramoto model does not take the existence of
background noise into account. The dynamics of interacting oscillators induced by UlDM with random noise is governed by
\begin{align}
\label{eq:noiseKura}
\dot{\theta}_{i} = \omega_{i} + \eta_{i} + g_{\omega} \frac{m_{\phi} \, c^{2}}{\hbar}\,r\, \text{sin}(\Psi - \theta_{i})\,,
\end{align}
where $\eta_{i}(t)$ stands for Gaussian white noise with the corresponding averaged correlations \cite{Sakaguchi:1988skh}
\begin{align}
\label{eq:noise}
\left\langle \eta_{i}(t)\right\rangle = 0,\,\,\,\,\,\, \left\langle \eta_{i}(t_{1})\, \eta_{j}(t_{2})\right\rangle = 2\,D\,\delta_{ij}\,\delta(t_{1}-t_{2})\,.
\end{align}

Here, the noise strength is denoted by $D$ with positive definite value $D \geq 0$. This induces the
Brownian motion of phase difference between the existing oscillators and impedes the synchronization
process. In this case, the critical coupling $g_{C}$ becomes\footnote{The critical coupling in the
presence of noise written in terms of original coupling is $K_{C} = \frac{2}{\int^{\infty}_{-\infty} \, d\omega\,G(\omega) \frac{D}{D^{2}+\omega^{2}}}$ \cite{Sakaguchi:1988skh}.}
\begin{align}
\label{eq:gCNoise}
g_{C} = \frac{2\hbar}{m_{\phi} c^{2}} \frac{1}{\int^{\infty}_{-\infty}\,d\omega\,G(\omega)\, \frac{D}{D^{2} + \omega^{2}}} \,.
\end{align}  

\begin{figure}
	\centering
	\includegraphics[width=0.9\textwidth]{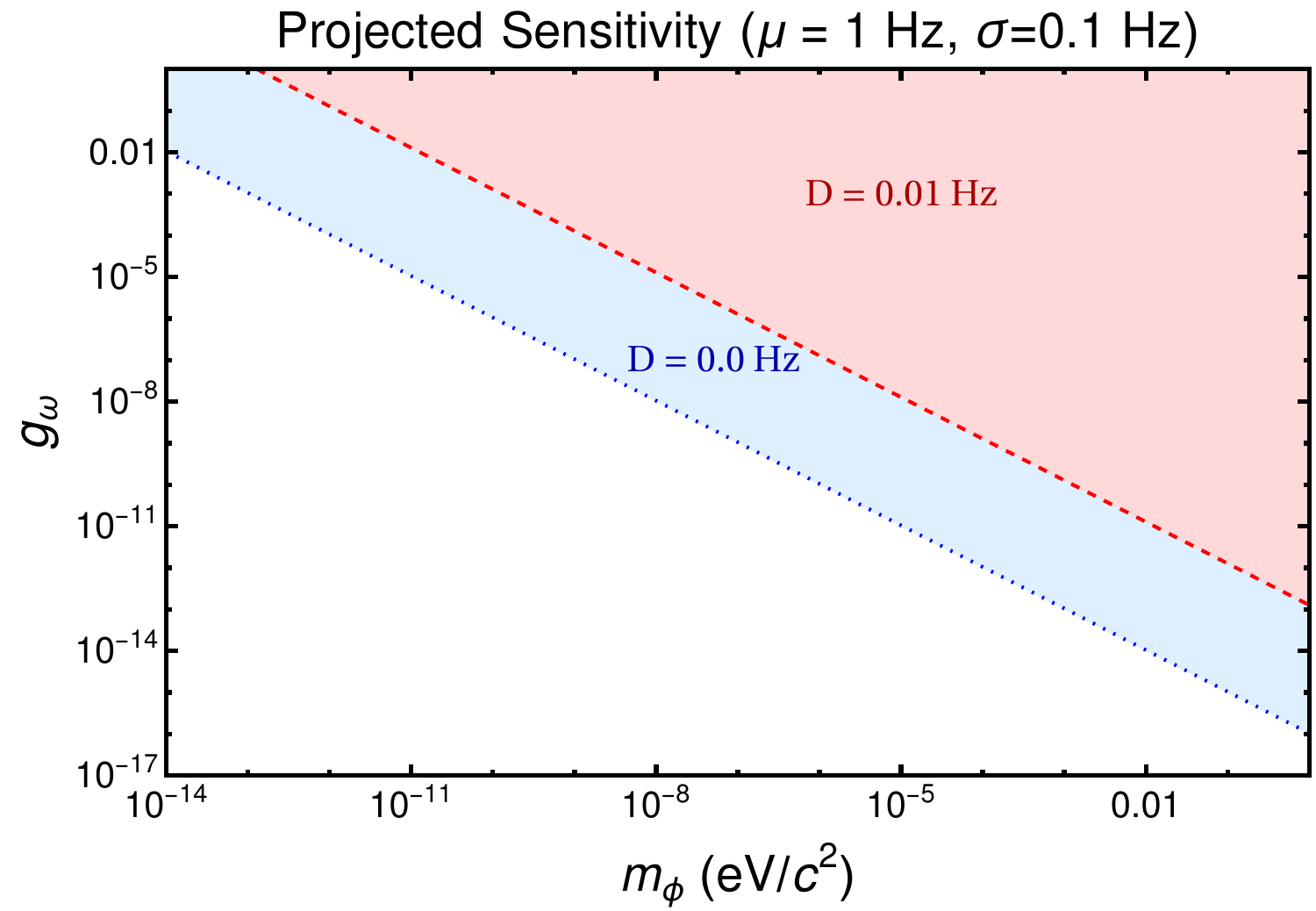}
	\caption{Projected sensitivites of the oscillators synchronized by UlDM. The blue region indicates the entrained oscillators without the noise while the red area displays the synchronization with Gaussian noise in the background.}
	\label{fig:sensitivity}
\end{figure}

However, if the interaction between oscillators and UlDM is strong enough or the noise is suppressed,
the synchronization would still take place. In Fig.\ref{fig:sensitivity}, we show the projected
sensitivity of particular distribution of oscillators with $\mu = 1$ Hz and $\sigma = 0.1$ Hz in $(m_{\phi},g_{\omega})$ plane. The blue dotted line indicates the critical coupling $g_{C}$ in Kuramoto
model without noise contribution while the light blue region above it indicates $g_{\omega} > g_{C}$
where the synchronization occurs. Here, we see that for UlDM mass within the interval $10^{-14}\, \text{eV}/c^{2} \leq m_{\phi} \leq 1\, \text{eV}/c^{2}$, synchronization occurs when the coupling lies
within $10^{-2} \geq g_{\omega} \geq 10^{-16}$. On the other hand, for noise strength $D = 0.01$ Hz \cite{Sakaguchi:1988skh},
the critical coupling is shifted to larger value $10 \geq g_{\omega} \geq 10^{-13}$ for the same UlDM
mass range as depicted by red dashed line. It becomes apparent that the oscillators would be moving in
unison when their coupling is greater than the new value of $g_{C}$ as shown in light red area.

\section{Summary and Conclusion}
\label{sec:Summary}

Collective behaviour is observed ubiquitously in nature.  We study the possibility that
ultralight dark matter would induce the synchronization behaviour on a given distribution of
oscillators randomly evolve according to their natural frequencies. For oscillators population with
mean angular frequency $1$ Hz  and standard deviation $0.1$ Hz, we demonstrate that UlDM with mass $10^{-14}\, \text{eV}/c^{2} \leq m_{\phi} \leq 1 \text{eV}/c^{2}$ and coupling larger than $10 \geq g_{\omega} \geq 10^{-13}$ would induce the synchronization on such oscillators, provided the Gaussian noise
strength in the background is lower than 0.01 Hz. 
By lowering both the standard deviation of the oscillators distribution as well as reducing the noise,
one can achieve better sensitivity. This opens a new venue to probe ultralight dark matter
complementary to those of the existing experiments.

\section*{Acknowledgment}
\label{sec:Acknowledgment}
This work was supported by the National Science and Technology Council (NSTC) of Taiwan under Grant No.NSTC 113-2811-M-003-019.

\end{document}